# Superstripes in the Low Energy Physics of Complex Quantum Matter at the Mesoscale

Antonio Bianconi

*Rome International Center for Materials Science Superstripes RICMASS,*
*Via dei Sabelli 119/A, 00185 Rome, Italy Email:* **antonio.bianconi@ricmass.eu**

Quantum physics in the XX century was proposed to understand the phenomenology of atomic world at short length scale (below one nanometer) and it was developed to study nuclear and subnuclear world at the lowest possible spatial scale and at the highest possible energy. Today in the XXI century the hot topic is to understand the mesoscale world extending in the range between 1 nanometer and 100 microns where the energy range of interactions is between 5 meV and 250 meV. From these studies a new "**Low Energy Physics**" (LEP) of *many body collective quantum phenomena* is emerging. The international conference "Superstripes - 2014", the last of the series of Superstripes conferences has been a key event for the scientists active in this new low energy physics. The focus has been on a) the competition between CDW and multigap superconductivity, and on b) the nanoscale phase separation in all high temperature superconductors. The conferences of the superstripes series are now the yearly major international event in the field of quantum phenomena in the complex quantum matter at the mesoscale.

The most relevant advances on our understanding of the spatial and time fluctuations in heterostructures at atomic limit have been presented at the Superstripes – 2014 conference, held in Erice, Italy on July 25-31, 2014. A new low energy (in the range between 5 and 250 meV) physics focusing in the mesoscale (extending in the range between 10 and 100 microns) world is emerging. The intricate mesoscale world driven by quantum critical charge, orbital and lattice fluctuations is now becoming the main topic of the Superstripes conference series.

This field started in 1992 driven by the interest to understand the nanoscale phase separation of both "electronic matter" and "lattice matter" in cuprate perovskites [1]. Followed in 1996 by the series of Stripes conferences [2] followed in 2008 by the present series of Superstripes conferences [3]. While the majority of the scientific community working in high temperature physics was accepting for 25 years the dogma of a uniform $CuO_2$ plane and a single electronic component, the small scientific community of the stripes conferences proposed that an intrinsic non trivial inhomogeneity, with different electronic components and anomalous local lattice fluctuations, was a key feature for the emergence of high temperature superconductivity [3-6]. The complex scenario object of discussions was characterized by a nanoscale phase separation with the competition of different phases in nanoscale domains: i) short range spin density wave SDW puddles; ii) short range incommensurate charge density wave ICD puddles, and iii) superconducting low dimension domains. This new scenario, called "superstripes" [7,8] was similar to emulsions in soft matter, where multiple phases at small length scales coexist. The superstripes scenario is today well supported by new experimental methods. The nanoscale phase separation takes place in the mesoscopic world intermediate between the atomic scale (for d less than 0.5nm) from the macroscopic world (for d larger than 100 microns). All experiments show that the essential interactions are in the Low Energy range which is the same range of many body interactions determining the emergence of life in the cell.

In material science the new low energy physics is determining advances in new nanotechnologies controlling the structure and textures in the mesoscopic scale.

Several research groups have reported at Superstripes-2014 advances on charge density wave physics [9-11]. A new type of light, synchrotron radiation (SR), to investigate the mesoscale matter organization became available worldwide only in this last 20 years. The invention of the synchrotron radiation source, the storage ring was made in the 60's, in Frascati by Bruno Touschek and the first large synchrotron radiation facility were developed at Stanford and Frascati in the 70's. But it was only in these last 20 years that synchrotron radiation was available worldwide and it has allowed to image and control Magnetic, Lattice, and Charge complexity in the mesoscale. The new methodologies probe multiscale spatial and temporal complexity in the low energy range 25-250 meV, opening new perspectives for low energy physics. These results are opening new roadmaps toward the design of new room temperature superconductors and new electronic and magnetic functional materials. The XANES and EXAFS methods [12,13] have been applied to investigate the fast local lattice fluctuations of the Cu-O bond [14-18] showing that nanoscale phase separation is an essential term in the physics of transition metal oxides showing emergent collective quantum phenomena. These results support the idea of Alex Müller where the electron lattice dynamical Jahn-Teller interaction is a key ingredient for the emergence of high temperature superconductivity (HTS). This has been neglected by the majority of the proposed theoretical mechanisms for HTS in cuprates and iron based superconductors for many years. At superstripes 2014 advanced SR methods, focusing on X-ray diffraction show the key role of local lattice distortions in high temperature superconductors and related materials [19-22].

At the Superstripes 2014 conference complex mesoscopic textures in transition metal oxides, graphene and silicene have been discussed and the control of the mesoscopic world for the emergence of high temperature superconductivity and electronic functionalities have been presented [23-25]

Advanced theories describing the complex multiphase physics driving the complex nanoscale phase separation have been discussed [26-30]. There is a consensus that the high temperature superconductivity emerges in a landscape where nanoscale ICDW puddles of polaronic incommensurate charge density wave (ICDW) with the associated incommensurate periodic lattice distortions (PLD) in the $CuO_2$ atomic layer coexist with superconducting filamentary domains. The scenario where spin fluctuations mediate high temperature superconductivy has been discussed by Plakida [31] and Jalborg [32] has presented the minimal requirements of band structure electronic features for superconductivity.

The physics of the unconventional superconducting phases made of multi-condensates [33] as been object of many reports [34-37] and of a school on solid state physics held in the previous week.

In conclusion there is agreement that 1) the material architecture of all high temperature superconductors is made of stacks of multiple atomic oxide layers where the active copper-oxide, iron or boron atomic planes are separated by spacer layers, 2) in the active planes multiple short range puddles of different electronic phases compete and cooperate 3) high temperature superconductivity is a particular case of multi-gap superconductivity.

**References**

1. K. A. Müller, G. Benedek "Phase Separation in Cuprate Superconductors" Proceedings of the Workshop on Phase Separation in Cuprate Superconductors, Erice, Erice, Italy, 6-12 May 1992 (World Scientific, 1993), URL http://www.worldcat.org/isbn/9789810212742.

2. A. Bianconi, N. L. Saini "Stripes and Related Phenomena" (Proceedings of the Stripes 1998 conference Rome, Italy) Springer, Selected topics in superconductivity 2000), ISBN 978-0-306-47100-1

3. A. Bianconi, Journal of Superconductivity and Novel Magnetism 24, 1117 (2011) URL http://dx.doi.org/10.1007/s10948-011-1142-4.

4. V. Kresin, Y. Ovchinnikov, S. Wolf Physics Reports 431, 231 (2006) doi:10.1016/j.physrep.2006.05.006 URL http://www.sciencedirect.com/science/article/pii/S0370157306001633

5. K. A. Müller, Journal of Superconductivity and Novel Magnetism 27, 2163-2179 (2014). URL http://dx.doi.org/10.1007/s10948-014-2751-5.

6. J. C., Phillips, Journal of Superconductivity and Novel Magnetism 27, 345-347 (2014). URL http://dx.doi.org/10.1007/s10948-013-2308-z

7. A. Bianconi Nature Phys. 9, 536-537 (2013), doi:10.1038/nphys2738 URL http://dx.doi.org/10.1038/nphys2738.

8. A. Bianconi, International Journal of Modern Physics B 14, 3289 (2000), doi:10.1142/S0217979200003769 URL http://dx.doi.org/10.1142/S0217979200003769.

9. S. Brazovskii, Journal of Superconductivity and Novel Magnetism (2014), pp. 1-5, doi:10.1007/s10948-014-2917-1 URL http://dx.doi.org/10.1007/s10948-014-2917-1.

10. T. Yi, Bravo, N. Kirova, and S. Brazovskii, Journal of Superconductivity and Novel Magnetism pp. 1-5, (2014) doi:10.1007/s10948-014-2916-2 URL http://dx.doi.org/10.1007/s10948-014-2916-2.

11. M. N. Gastiasoro and B. M. Andersen, Journal of Superconductivity and Novel Magnetism pp. 1-4, (2014) doi:10.1007/s10948-014-2908-2 URL http://dx.doi.org/10.1007/s10948-014-2908-2.

12. A. Bianconi, S. Doniach, and D. Lublin, Chemical Physics Letters 59, 121 (1978), doi:10.1016/0009-2614(78)85629-2 URL http://dx.doi.org/10.1016/0009-2614(78)85629-2.

13. J. Garcia, A. Bianconi, M. Benfatto, and C. R. Natoli, Le Journal de Physique Colloques 47, C8 (1986), URL http://dx.doi.org/10.1051/jphyscol:1986807.

14. A. Bianconi, M. Missori, H. Oyanagi, H. Yamaguchi, Y. Nishiara, and S. Della Longa, EPL (Europhysics Letters) 31, 411 (1995), doi:10.1209/0295-5075/31/7/012 URL http://iopscience.iop.org/0295-5075/31/7/012.

15. S. D., Conradson J. J., Mustre De Leon A. R. Bishop *J. Supercond.* 10, 329. (1997)

URL http://link.springer.com/article/10.1007/BF02765713#page-1

16. J. Mustre de Leon, M. Acosta-Alejandro, S. D. Conradson, and A. R. Bishop. Journal of Synchrotron Radiation 12, 193 (2005).
    doi:10.1107/S0909049505000063

17. J. Mustre de León, M. Acosta-Alejandro, S. D. Conradson, and A. R. Bishop Journal of Physics and Chemistry of Solids 69, 2288 (2008).
    doi: 10.1016/j.jpcs.2008.04.024

18. García Saravia Ortíz de Montellano and J. Mustre de León, Journal of Superconductivity and Novel Magnetism pp. 1-5, (2014), doi:10.1007/s10948-014-2913-5, URL http://dx.doi.org/10.1007/s10948-014-2913-5.

19. S. Shylin, V. Ksenofontov, S. Medvedev, V. Tsurkan, and C. Felser, Journal of Superconductivity and Novel Magnetism (2014) doi:10.1007/s10948-014-2912-6, URL http://dx.doi.org/10.1007/s10948-014-2912-6.

20. A. Athauda, J. Yang, B. Li, Y. Mizuguchi, S. Lee, and D. Louca, Journal of Superconductivity and Novel Magnetism (2014) doi:10.1007/s10948-014-2918-0 URL http://dx.doi.org/10.1007/s10948-014-2918-0.

21. A. Ricci, Journal of Superconductivity and Novel Magnetism (2014), doi:10.1007/s10948-014-2907-3 URL http://dx.doi.org/10.1007/s10948-014-2907-3.

22. A. Dusza, A. Lucarelli, J. H. Chu, I. R. Fisher, and L. Degiorgi, Journal of Superconductivity and Novel Magnetism (2014). doi:10.1007/s10948-014-2901-9 URL http://dx.doi.org/10.1007/s10948-014-2901-9.

23. N. Poccia, A. Ricci, F. Coneri, M. Stehno, G. Campi, N. Demitri, G. Bais, Wang, and H. Hilgenkamp, Journal of Superconductivity and Novel Magnetism (2014), doi:10.1007/s10948-014-2902-8, URL http://dx.doi.org/10.1007/s10948-014-2902-8.

24. M. Ezawa, Journal of Superconductivity and Novel Magnetism (2014). doi:10.1007/s10948-014-2900-x (2014), URL http://dx.doi.org/10.1007/s10948-014-2900-x.

25. N. Bovenzi, F. Finocchiaro, N. Scopigno, D. Bucheli, S. Caprara, G. Seibold, and M. Grilli, Journal of Superconductivity and Novel Magnetism (2014). doi:10.1007/s10948-014-2903-7 URL http://dx.doi.org/10.1007/s10948-014-2903-7.

26. G. Campi, D. Innocenti, and A. Bianconi, Journal of Superconductivity and Novel Magnetism pp. 1-9 (2015) URL http://dx.doi.org/10.1007/s10948-015-2955-3.

27. E. W. Carlson, S. Liu, B. Phillabaum, and K. A. Dahmen, Journal of Superconductivity and Novel Magnetism (2015), doi:10.1007/s10948-014-2898-0 URL http://dx.doi.org/10.1007/s10948-014-2898-0.
28. M. Saarela and F. V. Kusmartsev, Journal of Superconductivity and Novel Magnetism (2015), doi:10.1007/s10948-014-2915-3 URL http://dx.doi.org/10.1007/s10948-014-2915-3.
29. E. V. L. de Mello, Journal of Superconductivity and Novel Magnetism (2015), doi:10.1007/s10948-014-2899-z URL http://dx.doi.org/10.1007/s10948-014-2899-z.
30. K. Kapcia, Journal of Superconductivity and Novel Magnetism (2014), doi:10.1007/s10948-014-2906-4 URL http://dx.doi.org/10.1007/s10948-014-2906-4.
31. N. Plakida, Journal of Superconductivity and Novel Magnetism (2014). doi:10.1007/s10948-014-2911-7 URL http://dx.doi.org/10.1007/s10948-014-2911-7.
32. T. Jarlborg, Journal of Superconductivity and Novel Magnetism (2014). doi:10.1007/s10948-014-2897-1, URL http://dx.doi.org/10.1007/s10948-014-2897-1.
33. A. Bianconi, Journal of Superconductivity 18, 625 (2005), doi:10.1007/s10948-005-0047-5 URL http://dx.doi.org/10.1007/s10948-005-0047-5.
34. I. Chávez, M. Grether, and M. de Llano, Journal of Superconductivity and Novel Magnetism (2014). doi:10.1007/s10948-014-2904-6 URL http://dx.doi.org/10.1007/s10948-014-2904-6.
35. T. Yanagisawa, Journal of Superconductivity and Novel Magnetism (2014) doi:10.1007/s10948-014-2905-5 URL http://dx.doi.org/10.1007/s10948-014-2905-5.
36. H. Maebashi and Y. Takada, Journal of Superconductivity and Novel Magnetism (2014) doi:10.1007/s10948-014-2914-4 URL http://dx.doi.org/10.1007/s10948-014-2914-4.
37. M. D. Croitoru and A. I. Buzdin, Journal of Superconductivity and Novel Magnetism pp. 1-4, (2014). doi:10.1007/s10948-014-2910-8 URL http://dx.doi.org/10.1007/s10948-014-2910-8